\documentclass[epjCONF]{svjour}
\usepackage{graphics}
\usepackage[varg]{txfonts} 
\usepackage[latin1]{inputenc}
\session-title{Hot and Cold Baryonic Matter -- HCBM 2010}
\newcommand{\be}{\begin{eqnarray}}
\newcommand{\ee}{\end{eqnarray}}
\newcommand{\non}{\nonumber \\}
\newcommand{\ave}[1]{\langle #1 \rangle}

\begin{document}
\title{Facets of the QCD Phase-Diagram}
\author{Volker Koch\inst{1}\fnmsep\thanks{\email{vkoch@lbl.gov}} \and
  Adam Bzdak\inst{1}\fnmsep\thanks{\email{ABzdak@lbl.gov}}  \and
  Jinfeng Liao\inst{2} \fnmsep\thanks{\email{JLiao@bnl.gov}}  }
\institute{Lawrence Berkeley National Laboratory\\
Berkeley, CA 94720, USA
 \and 
Brookhaven National Laboratory\\
Upton, NY 11796, USA}
\abstract{
In this contribution we will discuss two aspects of the matter created
in ultra-relativistic heavy ion collisions. First we will
attempt to define a universal measure for the fluidity of a substance,
which will allow a correct comparison between the fluidity of a Quark
Gluon Plasma and any well known substance. Second we will
discuss current measurements of particle correlations and their
implication for possible local parity violation.
} 
\maketitle
\section{Introduction}
\label{intro}
Experiments at the Relativistic  Heavy Ion Collider (RHIC) have
revealed quite a number of interesting and surprising results. Among
them is the rather strong elliptic flow \cite{Voloshin:2008dg}, which
seems to suggest almost ideal fluid-dynamic expansion of the system
created in these collisions (see e.g \cite{Heinz:2009xj,Teaney:2009qa}
for a recent review). This observation has lead to the conjecture that
the matter created   
in these collisions is strongly interacting, with nearly ideal
fluidity and the phrase ``perfect fluid'' has been coined to describe
the matter at RHIC. This was mostly inspired by the observation that a
large class of gauge theories exhibit a lower bound on the ratio of
shear viscosity over entropy-density, $\eta/s\geq 1/(4\pi)$, in the
limit of very large coupling \cite{Policastro:2001yc}. And indeed it
seems that the ratio of $\eta/s\geq 1/(4\pi)$ appears to be a lower
limit to all known substances \cite{Csernai:2006zz,Lacey:2006bc}
including quantum liquids and cold Fermi gases \cite{Schafer:2009dj}. 
However, it is not obvious to which extent the ratio $\eta/s$ controls
or defines the fluidity of a substance. It certainly does not enter
naturally in the non-relativistic Navier-Stokes equation. To clarify
this situation, in the first part of this contribution we will discuss 
how one can define a measure of fluidity which is equally applicable
for both relativistic fluids, such as the Quark Gluon Plasma (QGP) as
well as non-relativistic fluids, such as water. Details can be found
in \cite{Liao:2009gb}. 

Another potentially interesting aspect of heavy ion collisions is that
they may be utilized to detect local parity violation due to the non-trivial
topological properties of the strong interaction, QCD. The
suggestions by Kharzeev and collaborators that the topological 
sphaleron transitions together with a chirally restored phase created
in heavy ion collisions could result in the so-called Chiral Magnetic
Effect (CME) \cite{Kharzeev:2007jp}. The Chiral Magnetic Effect
predicts that in the presence of the strong external (electrodynamic)
magnetic field  at the early stage after a (non-central) collision
sphaleron transitions induce a separation of charges along the
direction of the magnetic field. Of course, sphaleron and
anti-sphaleron transitions are equally likely, and, therefore, the
event-averaged charge separation will vanish. However, if present,
the event-by-event charge separation should be observable in a
suitable correlation measurement. Such a measurement has been proposed
by Voloshin in \cite{Voloshin:2004vk} and recently carried out by the
STAR collaboration \cite{Star:2009uh,Star:2009txa}. Here we will discuss to which extend the STAR
measurement is indeed sensitive to the CME. Details can be found in  \cite{Bzdak:2009fc,Liao:2010nv,Bzdak:2010fd}

\section{A universal measure for fluidity}
\label{sec:1}
Let us start this section by reminding ourselves about the
non-relativistic Navier-Stokes equation \cite{Landau_Hydro} 
(in the absence of bulk viscosity) 
 \begin{eqnarray}
[\partial_t + \vec v \cdot \vec \bigtriangledown] \vec v = -
\frac{\vec \bigtriangledown\, p}{\rho} + \frac{\eta}{\rho}
\vec{\bigtriangledown_j \Sigma^{j i}}
\end{eqnarray}
with the non-relativistic shear tensor $\Sigma_{ji}=\partial_j v_i
+\partial_i v_j-\frac{2}{3} \delta_{ji}\vec
\bigtriangledown\cdot\vec v$. From this expression, one already sees
that the dissipative term is controlled by the ratio of the shear
viscosity $\eta$ over the mass-density $\rho$, commonly referred to as
the kinematic viscosity \cite{Landau_Hydro} 
\be
\nu \equiv \frac{\eta}{\rho}
\ee
This is similar to classical mechanics where the ratio of friction
over inertia  and not the friction by itself controls the dynamics. And
indeed, while the shear viscosity of water is a hundred times larger
than that of air, its kinematic viscosity is a factor fifteen smaller,
supporting our everyday notion that water is a better fluid than air.
We further note the absence of a term containing the widely discussed ratio
$\eta/s$ in the non-relativistic Navier-Stokes equation. Hence it is
rather unlikely that this ratio can serve a measure for the fluidity
of non-relativistic fluids, such as water. Considering the corresponding relativistic
Navier-Stokes equation \cite{Landau_Hydro},
\begin{eqnarray}
\gamma^2  [\partial_t + \vec v \cdot \vec \bigtriangledown] \vec v
=&& - \frac{1}{w/c^2} [\vec \bigtriangledown\, p + \frac{\vec
v}{c}
\partial_0 p] \nonumber \\
&&  + \frac{\eta}{w/c^2} \vec{\partial_\nu \Sigma^{\nu i} } \,\,
\end{eqnarray}
with the relativistic shear tensor $\Sigma_{\mu\nu}=c\,
[\partial_\mu u_\nu + \partial_\nu u_\mu -(u\cdot
\partial)u_\mu u_\nu + \frac{2}{3}(u_\mu u_\nu - g_{\mu \nu})(\partial\cdot
u)]$, $\gamma=1/\sqrt{1-v^2/c^2}$, $u_\mu=\gamma(1,\vec v/c)$, and
$\partial_0 =\frac{1}{c}
\partial_t $. 
The essential difference from the non-relativistic version is that the
inertia is now given by the enthalpy-density $w$ instead of the
mass-density $\rho$. Thus the generalized form for the kinematic
viscosity is
\be
\nu = \frac{\eta}{w}
\ee
The enthalpy-density is given by
\be
w=\epsilon+p = Ts + \mu n
\ee
where $n$ is the particle density.
In the non-relativistic limit, $T \ll \mu$ so that $\mu \rightarrow m$ and
\be
w &\stackrel{T\ll\mu}{\longrightarrow}& m n =\rho, \non
\nu &\stackrel{T\ll\mu}{\longrightarrow}& \frac{\eta}{\rho},
\ee
recovering the non-relativistic Navier-Stokes equation. In the
relativistic limit, $T \gg \mu$, on the other hand
\be 
w &\stackrel{T\gg\mu}{\longrightarrow}& Ts, \non
\nu&\stackrel{T\gg\mu}{\longrightarrow}& \frac{\eta}{Ts}
\ee
Thus, in a sense the ratio $\eta/s$ is measure of the kinematic
viscosity for {\em relativistic} systems with small chemical
potential but certainly not for systems
like water, where the enthalpy-density is dominated by the mass-density. 

To define a more precise measure of fluidity, let us analyze the
propagation of sound modes. In the linearized regime, the dispersion
relation for a sound mode is given by
\begin{eqnarray} \label{eqn_sound_dispersion}
\omega= c_s\, k - \frac{i}{2}\, k^2\times  \left\{\begin{array}{c}
\frac{\frac{4}{3}\eta}{w/c^2}\, ,\quad
\mathrm{R\,\, fluid} \\
\frac{\frac{4}{3}\eta}{\rho}\, ,\quad \mathrm{NR\,\, fluid}
\end{array}\right\}
\end{eqnarray}
for a relativistic (R) fluid and non-relativistic (NR) fluid,
respectively. We note in passing, that the imaginary part, i.e. the dissipation, is
proportional to the {\em square} of the wave-number $k$ which reflects
the fact that fluid-dynamics works always in the long wavelength
limit. By requiring that the ratio of the imaginary over the real part
of the sound frequency is small, we ensure that the sound mode
propagates well. Or in other word the condition
\be
|\frac{{\mathcal Im}\omega}{{\mathcal Re}\omega}| \ll 1
\ee
results in a length-scale $L_{\eta}$ 
\be
L_{\eta} & \equiv & \left\{\begin{array}{c} \frac{\eta}{(w/c^2)\,
c_s}\, ,\quad
\mathrm{R\,\, fluid} \\
\frac{\eta}{\rho\, c_s}\, ,\quad \mathrm{NR\,\, fluid}
\end{array}\right\}
\ee
which measures the {\em minimum}
wavelength for which sound propagates with little damping while  for shorter
wavelengths dissipation becomes important. For a more precise
discussion we refer to \cite{Liao:2009gb}. As shown in
\cite{Liao:2009gb}, in the limit of dilute gases, this length-scale
$L_\eta$ is directly related to the mean free path, $L_\eta\sim \lambda_{mfp}$, which can be
defined for such a system. However, $L_\eta$ is not restricted to
dilute systems and can be extracted for any system, as it is defined
in terms of physical quantities, which do not required special conditions
such as a dilute gas. 

The final step is to relate $L_\eta$ to a length-scale {\em intrinsic}
to the system. This is necessary in order to be able to compare
system at vastly different length scales such as water, interstellar
dust and the QGP. The most obvious scale is the inter-particle
distance $d=n^{1/3}$, which is well defined for non-relativistic
systems. For relativistic systems made out of quasi-particles one can
extract the inter-particle distance from the entropy density $d \sim
s^{1/3}$. If no quasi-particle picture applies the typical
length-scale of a energy-density correlation function may serve the
purpose, which would also apply for other systems \cite{Liao:2009gb}. 

\begin{figure}
\resizebox{0.8\columnwidth}{!}{
\includegraphics{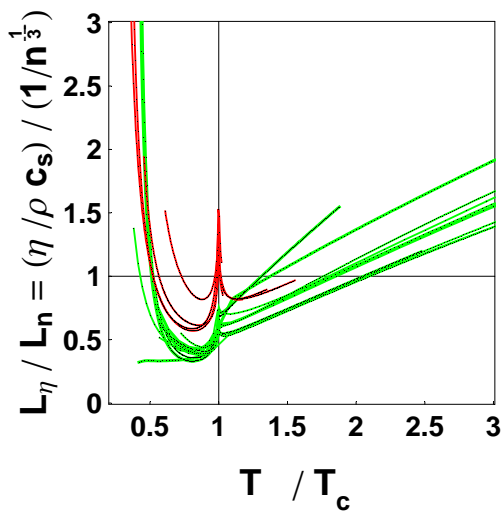} }
\caption {Fluidity measure ${\mathcal F}=
{L_\eta}/{L_n}$ versus $T/T_c$ for fifteen different substances 
($H_2$, $^4 He$, $H_2 O$, $D_2 O$, $Ne$, $N_2$, $O_2$,
$Ar$, $CO_2$, $Kr$, $Xe$, $C_4 H_{10}$, $C_8 H_{18}$, $C_{12}H_{26}$, $C_4 F_8$)
at
fixed critical pressure $P=P_c$. The data are from
\cite{NIST_webbook}.
The
sharp peaks centered at $T_c$ are due to the fact that the speed of
sound vanishes at $T_c$.}
\label{fig:fluidty_all}       
\end{figure}

Thus we define the fluidity measure ${\cal F}$ as 
\begin{eqnarray}
{\mathcal F}\equiv \frac{L_\eta}{L_n} \quad .
\end{eqnarray}
with 
\be
L_n = \frac{1}{n^{\frac{1}{3}}}
\ee
The resulting values for ${\cal F}$ for a large variety of substances
is shown in Fig.\ref{fig:fluidty_all}. 

Although the critical pressure and temperature as well as the molar mass of the
substances shown in Fig.~\ref{fig:fluidty_all} differ by several
orders of magnitude, their fluidity is quite similar, especially just
below the critical temperature. Thus, we may conclude that ``a good
fluid is a good fluid''. The spike at $T_c$ is due to the fact
that the velocity of sound, $c_s$, vanishes at the critical point. 

\begin{figure}
\resizebox{0.8\columnwidth}{!}{
\includegraphics{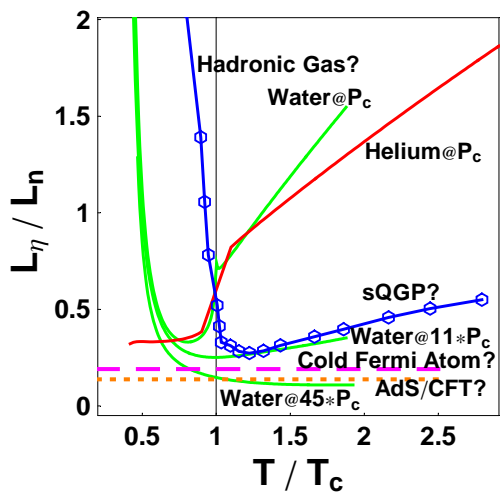} }
\caption{Comparison of the fluidity measure for various fluids
(see text for more details). The curves with question marks
indicate current estimates of the respective fluidity with
possible uncertainty, while the curves for Helium at $P_c$ and for
water at $P_c,11P_c,45P_c$ are from actual data.}
\label{fig:compare_fluid}       
\end{figure}

\begin{figure}
\resizebox{0.8\columnwidth}{!}{
\includegraphics{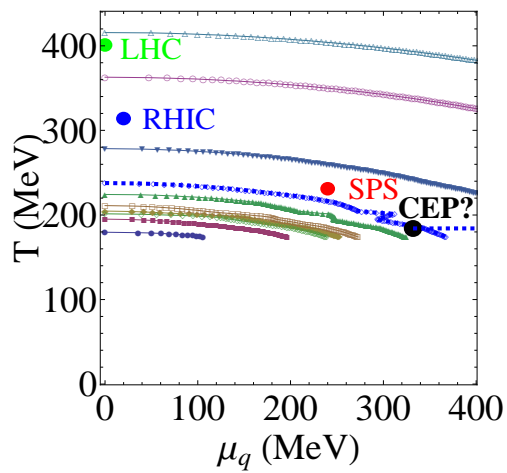} }
\caption{Schematic isobaric contours (i.e. with constant
pressure along each line) on the QCD $T-\mu$ phase diagram with
the filled black circle indicating a possible position of the {\em
hypothetical} Critical-End-Point (CEP); the dashed blue horizontal
short line stretching from the CEP to the right is included to
indicate the $T=T_{CEP}$ boundary (see text for more details). The
filled red, blue and green circles indicate estimates of the
initial $(T,\mu)$ reachable at SPS, RHIC and LHC, respectively.
}
\label{fig:super_qcd}       
\end{figure}

Since the fluidity measure seems to work rather well, it will be
interesting to see how the QGP compares with well known substances
such as water. This is shown
in Fig.~\ref{fig:compare_fluid}.
In order to estimate the fluidity measure ${\cal F}$ for the QGP, 
we have used a parametrization of the
viscosity $\eta$ by Hirano and Gyulassy \cite{Hirano:2005wx}. The
enthalpy-density $w=\epsilon+p$ and speed of sound $c_s$ are taken
from recent lattice results by Karsch et al \cite{Cheng:2007jq}
for 2+1 flavor QCD with $m_\pi\approx 220MeV$. As we mentioned
before, $L_n$ is estimated by $1/(s/4k_B)^{1/3}$ with the entropy
density also taken from \cite{Cheng:2007jq}. 
For the strongly coupled AdS/CFT system, the shear viscosity is
well known to be $\eta/s=1/(4\pi)$ \cite{Policastro:2001yc}. As we
also pointed out before, there is a short-range order at the
length scale $L_n\sim 1/T$ however the pre-factor is not
accurately determined. We simply use $L_n=1/T$ as an estimate.
This gives the fluidity ${\mathcal F}=\sqrt{3}/(4\pi)\approx
0.138$. For the Cold Fermi Atom gas we used available measurements for
its shear viscosity 
\cite{fermi_atom_viscosity} and the speed of sound
\cite{fermi_atom_sound}. For details see \cite{Liao:2009gb}. Finally
we also plot the fluidity measure for super-critical water at $P=11
P_C$ and $P= 45 P_c$. Surprisingly we find that the fluidity of
super-critical water is better or at least comparable with both the
QGP and the strongly coupled AdS/CFT system. Super-critical fluids are
substances with both the pressure and the temperature above the
critical point. They have the interesting property that they are
compressible but at the same time behave like fluids, as can be seen
from our plot. Since it is expected that the QCD phase-diagram
exhibits a critical point 
\cite{Stephanov:2004wx}, it is natural to speculate if the QGP may
be a super-critical fluid as well. This may very well be the case, as
demonstrated in Fig.~\ref{fig:super_qcd}, where we show the estimated
starting point for RHIC and LHC fluid-dynamic evolutions together with
equal pressure line in the QCD phase diagram. We also indicate a
possible position of the QCD critical point, noting that its existence
and precise location is still very much unknown.
What would be the consequences if indeed the QGP were a super-critical
fluid. Obviously, the system created at the LHC would then sit at even
higher temperature and pressure and thus should exhibit an even better
fluidity. And since the system at the LHC lives considerable longer
one would predict an even better description of the expansion dynamics
in terms of fluid-dynamics. To which extent this is borne out by the
recent measurement of elliptic flow at the LHC is questionable. The
dependence of $v_2$ on the transverse momentum is essentially the same
as observed for RHIC collisions, whereas a better fluidity would
predict a behavior more similar to ideal fluid-dynamics. However, before
one can draw any firm conclusions, the effect of jet-like correlations on
the elliptic flow measurements at higher transverse momentum needs to
be understood \cite{Liao:2009ni}.

Let us conclude this section with a few general remarks. First, the
ratio $\eta/s$ is not suitable to compare the fluidity of different
substances. It may serve as a measure for relativistic fluids far away
from a phase-transition, where the speed of sound is essentially a
constant.
Second, while $\eta/s$ may be small for a Quark Gluon Plasma, 
this does not mean that the sheer viscosity itself is small. Indeed,
using the conjectured lower limit of $\eta/s = 1/(4\pi)$ and inserting
the value for the entropy density of a QGP one arrives at
\be
\eta_{QGP}\simeq 2\times 10^9 \, {\rm Pa\, s} \simeq 10^{12} \eta_{Water}
\ee
which is probably the most ``sticky'' substance known. In comparison,
tar-pitch is estimated to have $\eta_{tar}\simeq 10^8 \, {\rm Pa\,s}$.
Thus the QGP flows only so well because it has such a large
inertia. 
The large value of the shear-viscosity itself could be
exposed if one would be able to drag a small object through the
QGP. In this case Stokes' formula for the force required to move an object
of radius $R$ at constant speed $u$ is
\be
F= 6 \pi R u \eta
\ee
Even if we rescaled the size of the sphere by the inter-particle distance,
$R_{QGP} \simeq 10^{-6}\,R_{Water}$, the force required to drag this sphere
through the QGP would still be six orders of magnitude larger than
that for water 
\be
F_{QGP}(R_{QGP}) \simeq 10^6 \, F_{Water}(R_{Water})
\ee
A sticky substance indeed! Expressing this drag force in terms of more
conventional units, 
\be
F_{QGP}\simeq 5 R u \frac{\rm GeV}{\rm fm^2}
\ee
so that a particle with radius of $R=0.5 fm$ and a velocity of $u=0.5$
would feel a drag force or energy loss of
$F=dE/dx \simeq 1.25 {\rm GeV/fm}$. To   
which extent this observation has any bearing on the observed energy
loss of heavy particle, remains to be seen.

\section{Local parity violation in heavy ion collisions?}
\label{sec:2}
\begin{figure}
\resizebox{0.75\columnwidth}{!}{
\includegraphics{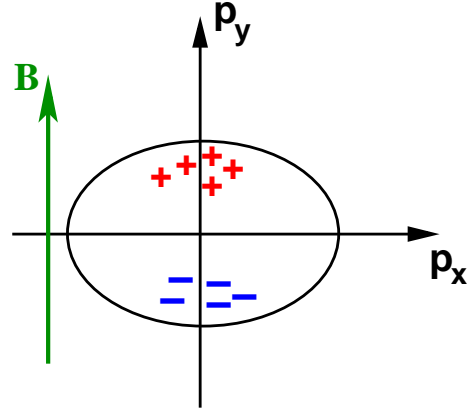} }
\caption{A schematic illustration of charge separation due to the
  Chiral Magnetic Effect in an heavy ion
  event. The reaction plane is aligned along the $p_x$-direction on this case.}
\label{fig:charge_separation}       
\end{figure}

As briefly discussed in the introduction, the Chiral Magnetic Effect (CME)
leads to the separation of charges along the  direction of the
magnetic field generated by the moving ions. This charge separation
can be viewed as a dipole in momentum space as depicted in
Fig.~\ref{fig:charge_separation}. In case of the CME, in a given event
the dipole vector will be either parallel or anti-parallel to the
magnetic field, depending on the presence of sphaleron- or
anti-sphaleron transitions in the reaction. Therefore,  the
expectation value of the momentum-space dipole-moment vanishes,
$\ave{\vec{d}}=0$, as does the expectation value of the parity-odd
scalar product with the  magnetic field, $\ave{\vec{B}\vec{d}}=0$.
However, since in case of charge separation $\ave{\vec{d}^2} \ne 0$ the presence
 of an event-by-event electric dipole may be observable in the
{\em variance} of a parity-odd operator, or equivalently, in
charge-dependent two-particle correlations. Of course, simple
statistical fluctuations also give rise to a finite $\ave{\vec{d}^2}$  
and suitable observables have to be devised which are not sensitive
to these statistical fluctuations (for a discussion see \cite{Liao:2010nv}).

One way to obtain information about the
presences of the CME is to study charge dependent two-particle
correlations with respect to the reaction plane, as proposed by
Voloshin \cite{Voloshin:2004vk}. He suggested to measure the following
three-particle correlation,
\be
\ave{\cos(\phi_i+\phi_j-2\phi_k)}
\ee
for same-charge pairs
($i,j=++/--$) and opposite-charge pairs ($i,j=+-$) with the third
particle, denoted by index $k$, having any charge. If the correlation
with the third particle $k$ is dominated by elliptic flow, then 
\be
\ave{\cos(\phi_i+\phi_j-2\phi_k)} = v_2 \,\ave{\cos(\phi_i+\phi_j-2\Psi_{R.P.})}
\ee
where $\Psi_{R.P.}$ is the angle of  the reaction plane, and $v_2$
denotes the strength of the elliptic flow. Working in a frame where
the reaction plane is along the x-axis,  $\Psi_{R.P.}=0$, we get
\be
\gamma \equiv  \frac{1}{v_2}  \ave{\cos(\phi_i+\phi_j-2\phi_k)} =
\ave{\cos(\phi_i+\phi_j)}
\ee

\begin{figure}
\resizebox{0.8\columnwidth}{!}{
\includegraphics{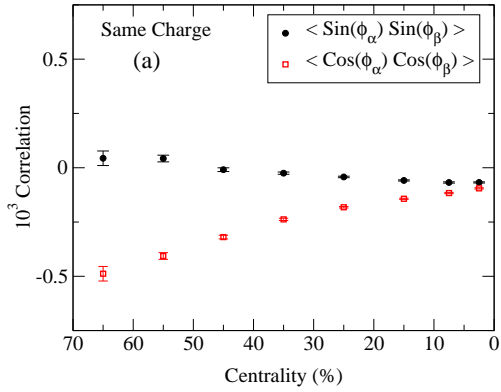} }
\caption{In-plane (red) and out-of-plane (black) correlations for
  same-charge pairs as measured by the STAR collaboration
  \cite{Star:2009uh,Star:2009txa}. For details see \cite{Bzdak:2009fc}.}
\label{fig:same}       
\end{figure}

\begin{figure}
\resizebox{0.8\columnwidth}{!}{
\includegraphics{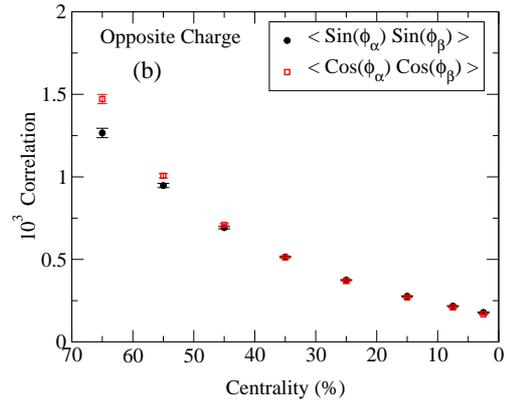} }
\caption{In-plane (red) and out-of-plane (black) correlations for
  opposite-charge pairs as measured by the STAR collaboration
  \cite{Star:2009uh,Star:2009txa}. For details see \cite{Bzdak:2009fc}.}

\label{fig:opposite}       
\end{figure}

\begin{figure}
\resizebox{0.8\columnwidth}{!}{
\includegraphics{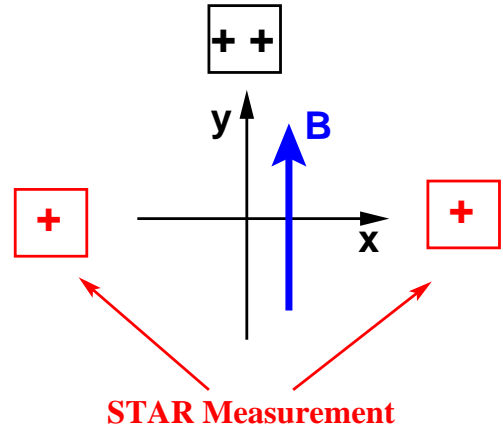} }
\caption{Schematic illustration of the actual STAR measurement (red)
  together with the predictions from the Chiral Magnetic Effect
  (black) for same-charge pairs.}
\label{fig:star_result}       
\end{figure}

The STAR collaboration 
has recently  measured this correlator  and indeed has
verified the above dependence of the elliptic flow. Before we discuss
the STAR measurement in detail, however, let us see what to expect for
this observable in case of the CME.
As can be seen from Fig.~\ref{fig:charge_separation}, the CME predicts
same-side out-of-plane correlations for same charges and back-to-back
out-of-plane correlations for opposite charges. This is best seen by
rewriting the correlator $\gamma$ as 
\be
\gamma = \ave{\cos(\phi_i+\phi_j)} &=&\ave{\cos(\phi_i)\cos(\phi_j)}
\non && - \ave{\sin(\phi_i)\sin(\phi_j)}
\ee
In this representation the first term, $\ave{\cos(\phi_i)\cos(\phi_j)}$,
measures the in-plane correlations while the second term, \linebreak
$\ave{\sin(\phi_i)\sin(\phi_j)}$, 
measures the out-of-plane
correlations. The CME predicts that same-charge pairs have either
both an angle of $\phi_i,\phi_j \simeq \pi/2$ or $\phi_i,\phi_j \simeq
3\pi/2$. In either case, $\sin(\phi_i)\sin(\phi_j)\simeq 1 $. For
  opposite charges,  $\phi_i \simeq \pi/2;\,\, \phi_j \simeq 3\pi/2$ or vice
  versa and $\sin(\phi_i)\sin(\phi_j)\simeq -1$. Hence the CME predicts 
\be
\gamma_{CME,\,same-charge} &<& 0\non
\gamma_{CME,\,opposite-charge} &>& 0,
\ee 
and indeed this is what the STAR measurement shows. So have we seen
the CME and thus local parity violation in an actual experiment? Not
quite yet, because there is an alternative scenario for which the
correlator $\gamma$ may be negative  for same charge pairs and
positive for opposite charge pairs: Suppose we have same-charge {\em in-plane}
back-back correlations, i.e. $\phi_i \simeq 0$ and $\phi_j \simeq \pi$ or vice
verse, and opposite-charge {\em in-plane} same-side correlations,
i.e.  $\phi_i,\phi_j \simeq 0$ or $\phi_i,\phi_j \simeq \pi$ we obtain the same
signs for $\gamma$ as above, but this time it is the
$\ave{\cos(\phi_i)\cos(\phi_j)}$ term which controls things. In other
words, the correlator $\gamma$ is not unique and we need another
observable to determine whether we are dealing with in-plane or
out-of-plane correlations. The obvious candidate is 
\be
\delta \equiv
\ave{\cos(\phi_i-\phi_j)}&=&\ave{\cos(\phi_i)\cos(\phi_j)}
\non &&+ \ave{\sin(\phi_i)\sin(\phi_j)}
\ee
which represents the {\em sum} of the in-plane
($\ave{\cos(\phi_i)\cos(\phi_j)}$) and  out-of plane
($\ave{\sin(\phi_i)\sin(\phi_j)}$) correlations. 
With both $\gamma$ and $\delta$ we can extract both in-plane and
out-of-plane correlations separately
\be
\ave{\cos(\phi_i)\cos(\phi_j)}&=&\frac{1}{2}(\delta + \gamma) \,\,\, {\rm
  (in-plane)} \non
\ave{\sin(\phi_i)\sin(\phi_j)}&=&\frac{1}{2}(\delta - \gamma) \,\,\, {\rm
  (out-of-plane)}
\ee
Fortunately, STAR has measured the correlator $\delta$ allowing for 
a decomposition of the in-plane and out-of-plane
correlations. Those are shown in Fig.~\ref{fig:same} for same-charge
pairs and in Fig.~\ref{fig:opposite} for opposite charge pairs.
The surprising result is that a) for same charge pairs the measured
out of-plane correlations are essentially zero, in contrast to the
predictions from the CME. Instead STAR observes an  
{\em in-plane} back-to-back correlation! This situation is illustrated
in Fig.~\ref{fig:star_result}. Opposite-charge pairs, on the other
hand seem to be equally correlated in both the in-plane and
out-of-plane direction. Obviously this is not quite in agreement with
the expectation from the CME. Especially the fact that the same-charge
pairs do not show any out-of-plane correlations for all centralities
is difficult to understand in the context of the CME
predictions. Naturally, there will be other effects contributing
to the correlators $\gamma$ and $\delta$, such as the coulomb
interaction, transverse-momentum conservation\cite{Bzdak:2010fd}, local charge
conservation \cite{Pratt:2010gy,Pratt:2010zn}, cluster-decays \cite{Wang:2009kd} etc. However,
it is difficult to imagine how for all centralities these ``background''
contributions conspire to perfectly cancel the
correlations expected from the CME.
One should note, however, that so far the measured correlation are not
understood in terms of conventional physics either, possibly because
many effects contribute, as indicated above. Furthermore, it would be
very useful to have a more differential information on the above
correlations. While STAR has extracted the rapidity and
transverse-momentum dependence of $\gamma$ this information is not yet
available for $\delta$. In addition, the value
for the above correlations in simple proton-proton collisions would
serve as an important reference point. 
It may also be useful to develop alternative observables
\cite{Liao:2010nv,Ajitanand:2010rc}. For example in \cite{Liao:2010nv} 
the direct extraction of the magnitude and direction with respect to
the reaction plane of the momentum-space dipole-moment has been proposed by introducing a charge-dependent Q-vector
analysis \cite{Voloshin:2008dg}. In \cite{Liao:2010nv} it was also
demonstrated that  simple
two-particle correlations may mimic the effect of an actual dipole,
and only the careful analysis of the distributions of both the
magnitude and the angle of the extracted dipole  was able to
distinguish between an explicit dipole and other correlations.

\section{Conclusions}
\label{sec:3}
In this contribution we have discussed two aspects of the physics of
dense matter. On the one hand we presented a universal measure for
the fluidity of any substance and we argued that the widely used measure
$\eta/s$ is not suitable for a comparison between non-relativistic and
relativistic fluids. Given our new measure, the fluidity of the QGP is
not any better than that of water. We further pointed out, that
super-critical fluids exhibit an exceptionally good fluidity and
speculated to which extent the QGP may be considered a super-critical
fluid. We also reminded ourselves that the shear viscosity of the QGP
is actually extremely large, and that it only flows so well because of
its high inertia, i.e. energy density. This large viscosity could be
revealed by dragging an object through the QGP and the resulting
drag-force turns out to be of the same magnitude as the 
energy loss extracted from the analysis of leading particle suppression.

In the second part we critically examined the STAR measurement of
charge dependent two and three particle correlations and their
relevance for local parity violation. We found that for same charge
pairs  STAR measures in-plane back-to-back correlations in
contradistinction to the prediction from the Chiral Magnetic effect,
which predicts out-of plane same side correlations. Therefore, the
jury on the existence of local parity violation in heavy ion collision
is still out.

\section{Acknowledgments}
This work was supported by the Director, 
Office of Science, Office of High Energy and Nuclear Physics, 
Division of Nuclear Physics, and by the Office of Basic Energy
Sciences, Division of Nuclear Sciences, of the U.S. Department of Energy 
under Contract No. DE-AC03-76SF00098 and Contract No. DOE Contract
No. DE-AC02-98CH10886 as well as by the Polish Ministry of
Science and Higher Education, grant No. N202 125437.

%
%
%
%

\end{document}